\documentstyle[12pt]{article}
\newcommand{\be}{\begin{equation}}
\newcommand{\ee}{\end{equation}}
\newcommand{\ba}{\begin{array}}
\newcommand{\ea}{\end{array}}
\newcommand{\bd}{\begin{displaymath}}
\newcommand{\ed}{\end{displaymath}}
\newcommand{\bt}{\begin{tabular}}
\newcommand{\et}{\end{tabular}}
\newcommand{\un}{\underline}
\newcommand{\bc}{\begin{center}}
\newcommand{\ec}{\end{center}}

\begin{document}

\large
\bc
 {\bf   ON THE NON-PERTURBATIVE PROPERTIES OF THE YANG-MILLS\\[3mm]
        VACUUM AND THE VACUUM ENERGY DENSITY    }
\ec
\bc
               V.G. KSENZOV
\ec
\vspace{5mm}
\bc
  {\it Institute of Theoretical and Experimental Physics,\\
    Moscow,Russia }

\ec
\vspace{12mm}

\bc
        A b s t r a c t
\ec
\vspace{8mm}

 The non-perturbative part of the vacuum energy density for static
conf\/igurations  in  pure $ SU(2) $ Y-M theory  is described.
The vacuum state is constructed.

\vspace{10cm}

\un{~~~~~~~~~~~~~~~~~~~~~~~~}

e-mail:ksenzov@vxitep.itep.ru

\newpage

\section{Introduction}

  The goal of the present paper  is to investigate  non-perturbative
dynamics in  the  pure Y-M theory which is the key to the calculation
of the vacuum condensates.

To understand the actual dynamics and the role of the non-perturbative
effects we should have an explicit form of the non-perturbative f\/ields.
As a rule the non-perturbative effects are associated with  classical
f\/ields characterized by topological charges.
To see how the non- perturbative
fluctuations generate  physical amplitudes we shall treat the
non-perturbative part of the vacuum expectation
value (v.e.v.) of the energy-momentum tensor
$\theta_{\mu \mu}$ in Y-M theory in four dimensions. In the case of theories
without dimensional parameters the
v.e.v. of $\theta_{\mu \mu}$ gives the vacuum energy density
$\varepsilon_{vac}$  and the characteristic mass scale thus def\/ining
properties of the
effective theory \cite {NS}. It has been  noticed that the connection
of non-perturbative
effects with  classical f\/ields  is not always straightforward. Indeed,
there is a set of models, such as $O(N)$ non-linear sigma models ($N>3$)
which do not have topological solutions, but do have
non-perturbative effects which have  been obtained
within perturbation theory \cite{NS, BL, RP}.
Besides, the physical vacuum has no topological charge $(Q=0)$
by def\/inition but we believe  its structure has a non-perturbative nature.

There  are  several methods of calculation of the
non-perturbative contributions into
v.e.v.'s of different correlations.
A  solution of the
problem in Y-M theory have been suggested by 't Hooft \cite{tH} and was a
subject of intense study. However, there are several  problems arising within
this approach  when instanton background is considered in the dilute-gas
approximation
\cite{CDG}. One of them is that contributions
to the integrals over the  instanton
size come from the region  of large sizes where the initial approximation is no
longer valid. Therefore, the exact magnitude of the non-perturbative effects
remains unknown.
Recently an alternative method has been proposed
in which the non-perturbative dynamics  has been investigated for the example
of the two dimensional $ O(N) $  non-linear sigma model in the
large  $N$ limit \cite{K,V}. In this model any instanton effects are absent.
Here we shall use the basic ideas which have been developed in these papers.
For this reason we recall the main features of the approach.

It was shown  that fluctuations describing the vacuum
properties of a theory are subject to  the requirements for potential energy
to be in the  minimum
and for conjugated canonical momenta to be zero. Such fluctuations are
not operators but a c-number function. For similar reason the
temporal component of the gauge f\/ield are c-number functions.

It is evident that constant f\/ields may obey these conditions. In such a case
the quantum fluctuations around the constant
background describe the perturbative
properties of the theory. If non-constant f\/ields satisfy these conditions
then the quantum fluctuations around the  non-constant  background describe
the non-perturbative properties of the theory.

The regularization procedure is essential for calculation  of vacuum
condensates. The regularization by separating the quantum f\/ields in
different points is used.To this end the point of the space in which
the regularization procedure has to be done should be replaced by a sphere
$ S^2 $ having
a small radius $ r $ which is set to zero at the very end of the calculation.
The quantum f\/ield is def\/ined on the  surface of the sphere.
From dimensional consideration
$ \varepsilon_{vac} $ has to be proportional to  $ 1/r^4 $,
therefore, naive  $ \varepsilon_{vac} $ goes to infinity when $r$ tends to zero.
In real fact the non-perturbative  value of $ \varepsilon_{vac} $ is f\/inite
owing to quantum effects.

The non-trivial classical f\/ields, mentioned above, are characterized by the
topological charge $ Q $. The energy funtional for static  non-trivial
configuration is defined as $ E=4{\pi}Q/g^2 $,\,where $g$ is
the coupling constant.
Obviously that the physical vacuum has no topological charge and may be
presented
as a sum of  classical configurations which  contribute
 to the topological charge with different signs.
However, there is an alternative possibility.
If the physical space has two boundaries, the topological charge
 may be equal to zero for non-trivial classical configurations.
Such situation emerges due to the suggested regularization method.
The point is
that  introduction  of the sphere of  small radius gives one more boundary
(another boundary is at large distance) and the topological charge is
given by the difference  of   contributions from the boundaries  and is
equal to zero for the inf\/initely small  radius \cite{R}\, say,
for the radius which is inverse  to the mass of the ultraviolet cutoff.

In the present paper we shall adop the basic ideas
which are obtained in sigma models
to investigate the non-perturbative dynamics and to  calculate the vacuum energy
density in Y-M theory.
In sect.2 the non-perturbative structure of vacuum state is
discussed in general form.
We describe the classical f\/ields
and discuss their properties in sect.3. In sect.4 we calculate the vacuum
energy density. It is necessary to stress
that the vacuum state is not obtained by solving the Schodinger equation.
We belive that the constructed  vacuum state is one of a coherent type.

\section{The non-perturbative structure of the vacuum state.}

  In the present section we suggest the method
of construting of the  non-perturbative  vacuum state using
perturbative vacuun state. Our basic idea can be better
explained in terms of the quantum mechanics.

Let us assume that $ \Psi(x) $  is the ground state of some
quantum system and $ T $ is the translation operator.
We know that the average value of the quantity $x+x_0=T^{-1}xT $ is determined by
the following expresion

$$
    \bar x=\int^{\infty}_{-\infty} dx\Psi^*(x)\,(x+x_0)\,\Psi(x)=
\int_{-\infty}^{\infty} dx\,\left(T\Psi\right)^*\,x\,\left(T\Psi\right)=x_0.
$$
This example shows we do not need to know, generally speaking, the explicit
ground-state $\Psi(x-x_0)$  to obtain the average value of $ \bar x\ne 0 $.
We can get it knowing only the translation operator and the parameter of the
transformation  $ x_0 $. Notice that $\Psi(x)$ and $\Psi(x-x_0)$  do not
satisfy  the requirement of orthogonality.
If the anologous  situation takes place in
quantum f\/ield theory then  having  the non-perturbative fluctuation
and some translation operator one can construct the vacuum
state and to calculate
vacuum condensate. However, contrary to quantum mechanics one has to know
some starting ground state in field theory because of v.e.v. is obtained
by another mean.It turns out that perturbative ground state may be selected
as the starting state. This situation facilitates the problem.

It was shown in the framework of the  sigma models in two dimensions that
this idea leads  to the correct result for the vacuum condensate of
$\theta_{\mu\mu} $ \cite{V}. Therefore we are about to discuss how
the idea is realized in quantum f\/ield theory.

Let us show how  the non-perturbative state is constructed. At the beginning the
non-perturbative vacuum f\/ields fluctuations are def\/ined
and then the translation operator and the non-perturbative vacuum state
is constructed.

Assume that $ L $ is the Lagrangian of some f\/ield system and $ h $ is
the Hamiltonian density. There is  a relation
$$
          L=\dot \phi\pi_0 - h,
$$
where $\phi $ is the f\/ield and $\pi_{\mu}=\delta L/\delta\partial_{\mu}\phi$
is the canonical momentum. Here group indexes are omitted.
The Hamiltonian is def\/ined as

$$         H=\int dx^{n-1}\,(\pi^2_{\mu} +V(\phi)).$$

Let us assume that the potential gets the minimum $V(\phi_v)=0$
when $\phi_v^2(x)=const$.
The finite energy condition is satisfied if
$ \pi_{\mu}=0,  V(\phi)=0 $  everywhere at large distances.
The solutions satisfying this condition  are a set
of  the vacuum f\/ields because we believe that  fields take their
vacuum value at large distances in theory without sources. It is believed
that vacuum is costructed in the same way at any point of the
physical space. Therefore the condition
defines the vacuum f\/ield in any point of the physical space except
a singular point.
The vacuum f\/ield can not be quantizated due to the
fact that their conjugated canonical momenta are equal to zero $ \pi_0=0 $ and
$\phi_v(x)$ is c-number function.

If the canonical momentum is $ \pi_{\mu}=\partial_{\mu}\phi$, then
the solutions of the condition are  trivial $\phi_v=const$. In this case
the theory is quantized around constant background and the quantum
fluctuations describe the perturbative properties of the theory  \cite{V}.

If the canonical momentum involves  a gauge f\/ield $A_{\mu}$,i.e.
$ \pi_{\mu}=\partial_{\mu}\phi+gA_{\mu}\phi $ ,then there are solutions of
the condition $\pi_{\mu}=o$ which are topologically
trivial $A_{\mu}=0, \phi_0=const $
 and  topologically non-trivial
 $A_{\mu}=1/gU\partial_{\mu}U^{-1}, \phi_v=U\phi_0 $ at large distances.
 U(x) is an arbitrary element of a gauge group and must be non-singular
 only on a bondary. The  f\/ields $ \phi_v(x)$ are c-number functions only
 if we quantize the theory over the non-constant background
 $\pi_0=\partial_o\phi_v + U\partial_{\mu}U^{-1}\phi_v=0$.
In the perturbation theory,
 when $A_{\mu}=0$ on the boundary and non-singular in all physical space,
 the canonical momentum $\pi_{\mu}=\partial_{\mu}\phi\ne0$ therefore the
 f\/ields $ \phi_v(x)$  are quantum f\/ields which describe
 the non-perturbative properties of the theory as will be argued in the paper.
As discussed in \cite{V} this is the reason why some results which are obtained
in the perturbative methods are related to the non-perturbative effects.

Let us suppose we quantize our theory  over the constant background.
If the constant is equal to zero then the vacuum of the  theory
may be described in a usual way as the eigenfunction of the
annihilation operator $ a_0 $ with zero eigenvalue of the
momentum  $ a_0|0>=0 $ \cite{IZ}  and v.e.v. of $\phi(x) $ is
$ <0|\phi(x)|0>=0 $. If one wants to have $ <0|\phi(x)|0>\ne0 $
then a  new vacuum state have to be constructed. The operator which
carries  out the  transformation $\phi \to v $ is well known to be
the operator of the canonical momentum. The translation operator is
$$
 T= \exp v\int dx\,\partial_0\phi =\exp v(a_0-a_0^+)
$$
and  the new vacuum state is written as $|v>=T|0> $.
In this case   the annihilation operator $ a_0 $ has  no
zero  eigenvalue  $a_0|v>=v|v>$ and v.e.v. of $\phi(x)$ is
$ <v|\phi(x)|v>=v$ \cite{IZ}.

If the vacuum fluctuation is  a function then the translation
operator can be written as
$$
   T=\exp\,\int dx^{n-1}\,(\phi_v(x)\pi_0\,+\,\partial_0\phi_v(x)\phi(x)).
$$
Here the last term in the integral  generates the translation
of the canonical momentum into $ \partial_0\phi_v(x) $.

Considering v.e.v. of some operator in the non-perturbative vacuum state
$  |\phi_v>=T|0>$  we obtain that all f\/ields operators are substituted
on the vacuum fluctuation  and  then   v.e.v.
have to   be calculated using the same method of regularization
as is given below.   Notice that the vacuum state
 $ T|0> $ is the total vacuum state which contains  perturbative
part $|0>$  and non-perturbative part  $ (T-1)|0> $ of the vacuum
state therefore there is no orthogonality between   the  perturbative
 and  the total vacuum states.

Here we recall that if there is a symmetry in the  theory then the symmetry
generator $Q $ annihilates the vacuum state $Q|0>=0$.
The condition of the symmetry
 breaking is the absence of the symmetry invariance of the vacuum state
$Q|0>\ne0$. In is necessary to stress that the symmetry generator is common
one but the vacuum states are different states. If we should like to
obtain the vacuum energy changes then  Hamiltonian have  to be  precisely
the same in the vacuum states.

The anologous  situation  takes place in this case.
By choosing as the  starting
state the vacuum state of the perturbative theory we have to use free
Hamiltonian of the theory and non-perturbative vacuum state to obtain
vacuum energy changes.

However there is a problem arising within this approach in Y-M theory.
The point is that free  Hamiltonian of the gauge f\/ields is not invariant
under  SU(2) local gauge transformation corresponding to Y-M theory.
Therefore the vacuum energy which is obtained by this the method
is gauge dependent.

It is known \cite{NS} that the vacuum energy may be
decomposed into non-perturbative and perturbative parts.  The pertubative
part is scale dependent. The non-perturbative part is scale
independent. Besids the non-perturbative part of the vacuum energy is
defined on the non-trivial topological f\/ields
because the physical local gauge
transformations which are trivial from the point of view of topology
can not break down the non-perturbative results. The gauge
transformations change only the perturbative part of the vacuum
energy which should be subtracted because it has no physical meaning.

Notice that the vacuum energy density is usually obtained as v.e.v. of the
trace of the energy-momentum tensor
which have to be defined in gauge non-invariant way
according to the above arguments.

 The translation operator in Y-M theory is def\/inefed as
$$
        T\exp\,\int dx^3 J^a_i\,E^a_i
$$
in the case of the static vacuum cof\/igurations. Using this vacuum state
we can obtain v.e.v. of $\theta_{\mu\mu}$ by the operator method.However
for pedagogical purpose the calculation of the vacuum energy density
in Y-M theory is carried out using functional integral.

\section{Classical vacuum in Yang-Mills theory}

The structure of the vacuum in the pure Y-M  theory was well studied. An
extensive reference list can be f\/ind in \cite{R}.
Here we are interested in static
properties of the vacuum. At the begining we will remind the formulation
given in paper \cite{P}. To set out notations we briefly recapitulate some
results relevant to our work.

The Hamiltonian of the classical Y-M theory in Euclidian space is
\be
\label{ham}
   H=\frac{1}{2}\int d^3 x\left( E^a_k\right)^2+\left( B^a_k\right)^2,
\ee
where $ E^a_k=F^a_{k4}$, $ B^a_k=1/2\varepsilon_{klm}F^a_{lm}$,
$ F^a_{ij}=\partial_i A^a_j-\partial_j A^a_i +g\varepsilon^{abc}A^a_i A^c_j $;
$ a,b,c$  are the group indices; $ k,l,m=1,2,3 $.
Let the gauge group be  $ SU\left(2\right) $ and
$\left(t^a\right)^{bc}=
\varepsilon^{abc} $
be  generators of the group $ SU\left(2\right) $
in the adjoint representation. In this case
 the f\/ields $ A^a_{\mu} $  are real. We also imply that the gauge
f\/ields are subject to the f\/irst class constraint and there is the gauge
f\/ixing condition  $ \partial_{\mu}A_{\mu}=0 $.

The gauge f\/ields which describe the classical properties of the vacuum
 must  satisfy the condition $ H=0 $ in the static case.
This, in its turn, yields
\be
\label{con}
             F^a_{4k}=F^a_{lm}=0,
\ee
i.e. the canonical momenta are equal to zero.
It follows that the f\/ields which describe the classical vacuum of the theory
are pure gauge.

The space components of the potential $ A $ are
\be\label{ga}
            A_i=\frac{1}{g}U(x)\partial_iU^{T}(x) ,
\ee
where $U(x)$  is an arbitrary element of the gauge group
which is taken independent of $ x_4 $ and $ U(x)U(x)^{T}=1 $.
Here  it is necessary to stress that the gauge f\/ield (\ref{ga}) may be
eliminated  by a gauge transformation only if  $  U(x) $ goes to zero
at large distances. However we shall also deal with
non-trivial topological   conf\/igurations therefore the function $ U(x) $
must  be non-singular on the  boundary and corresponds to
topological charge. As has already been mentioned,\,it is possible that
the total topological charge  and energy functional are equal to
zero even when there are non-trivial topological f\/ields.
Such possibility is discussed in the next section.
In this case $ A_i $ (\ref{ga}) can not be eliminated by gauge
transformations \cite{R}.

The components of the gauge field $ A_{4 } $ are usually chosen
equal to zero but, as shown in \cite{P}, $ A_{4}\ne0 $ def\/ines the
topological charge of the monopole and, as we shall see later,
other non-perturbative properties of the  theory in the static case.

The static conf\/igurations $ A^a_4 $ can be obtained from the
condition~(\ref{con})
\be
\label{fo}
            F^a_{4k}=\partial_k A^a_4+g \varepsilon^{abc} A^b_k A^c_4=0.
\ee

Multiplying this equation  by $ A^a_{4} $  and taking into
account antisymmetricity of the  tensor $ \varepsilon^{abc} $, we get
\be
\label{ca4}
         \partial_k\left(A^a_4\right)^2=0,i.e. \left(A^a_4\right)^2=const.
\ee

The constant in~(\ref{ca4}) is  arbitrary  but  it is important
that it may be chosen non-zero \cite{P}.
Although   the classical theory has no  dimensional parameter
we can introduce arbitrary quantity ${\mu}$ due to  the
condition (\ref{ca4}) and
define
\be
\label{mu}
                 A^a_4(x)={\mu}\Phi^a(x),
\ee
where $ \Phi^a $   is a dimensionless function which is subject to the condition
\be
\label{cf}
                \Phi^a(x)\Phi^a(x)=1.
\ee
It  follows from~(\ref{cf}) that the f\/ields $ \Phi^a(x) $  can be represented
as
\be
\label{uf}
                   \Phi(x)=U(x)\Phi_0,
\ee
where $ \Phi_0 $ is a constant vector in the colour space. As
follows from ~eq.(\ref{uf}) the shift of $\phi $ in $x$-space is
equivalent to the rotation of $\phi_0$ in the colour space.

The vector potential $ A^a_k $  may be expressed by a combination of the
f\/ields $ \Phi^a(x) $. Multiplying eq.(\ref{fo}) by $ \varepsilon^{ial}\Phi^l $
we obtain
\be
\label{ja}
                      J^i_k=g\left(\delta^{in} -\Phi^i\Phi^n\right)A^n_k,
\ee
where $ J^i_k=\varepsilon^{ial}\partial_k\Phi^a\Phi^l $.
Introducing $ P^{in}=\delta^{in} - \Phi^i\Phi^n $  then
$ P^{in}\Phi^n=0 $ and $ P^{in}P^{nl}=P^{il} $. Therefore multiplying (\ref{ja})
by $ P^{if} $ we get
\be
\label{p}
                P^{if}\left(J^f_k - g A^f_k\right)=0.
\ee
The solution of eq. (\ref{p}) is
\be
\label{a}
             A^a_k=\frac{1}{g}\left(J^a_k-n_k\Phi^a\right),
\ee
where $ n_k(x) $ is an arbitrary function.
The solution (\ref{a}) also satisfies the condition $ F_{ij}=0 $ (\ref{con})
if $ n_k(x) $ is a pure gauge in the group $ U(1) $, i.e.
$ \partial_in_j~-~\partial_jn_i~=~0 $.
The quantity $ n_i $ def\/ines the projection of  $ A^a_k $ onto $ \Phi^a(x) $.
Indeed, from  (\ref{a}) we get
\be
\label{ni}
                  n_k=-gA^a_k\Phi^a(x)
\ee

$ J^a_i $ is a conserved current
$ \left(\partial_iJ^a_i=0\right) $   associated with the global symmetry
$  SO(3) $ of the theory. The conservation of the current follows from the
relation
$$
        \partial^2\Phi^a=\Phi^a\left(\partial_i\Phi^a\right)^2,
$$
which can be obtained from eq.(\ref{cf}).

The gauge f\/ixing condition $ \partial_{\mu}A_{\mu}=\partial_kA_k=0 $
may be satisf\/ied by requiring the orthogonality of $ \Phi^a $ and
$ A^a_k $ or $ n_k=0 $, i.e. $ n_k $ is a gauge-fixing parameter.
In such a case the gauge potential (\ref{a})
 is not fixed by the Coulomb gauge. The property of the
Coulomb gauge was first discussed by V.N. Gribov \cite{G}.

Let us find  how the f\/ields  $  \Phi^a(x) $   are expressed in terms  of the
current $ J^a_k $. To do this we may use eq. (\ref{fo}). Substituting
(\ref{mu}) and(\ref{a}) into (\ref{fo}) we obtain
\be
\label{ph}
            \partial_k\Phi^a(x) +\varepsilon^{abc}J^b_k(x)\Phi^c(x)=0.
\ee
The solution of eq.(\ref{ph}) can be written as
\be
\label{pin}
      \Phi(x)=P \exp\left(-\int\limits^x_{x_0}d z_i\,J_i\right)\Phi(x_0)~.
\ee
where $ J_i=J^a_it^a $ and $ x_0 $ is an arbitrary point. As it is clear
from (\ref{pin}) and (\ref{cf}) the values of the  f\/ield  $  \Phi(x) $
in two different fixed points $ x_1 $
and $ x_2 $  are related to each other by the global group transformation and
any  invariant of the global transformation is independent of
choosing the point  $ x_0 $.
 One can see from (\ref{ga}) and(\ref{a})
that the current $ J_k $ is def\/inded as
\be
\label{gj}
                   J_k=U\partial_kU^{-1},
\ee
when $ n_k=0 $. Therefore the integral (\ref{pin}) does not depend on
the integration path if the function $ U(x) $ has no singularities.
However to have non-trivial topological effects we should also admit the
occurrence of gauge functions which have a singularity. Then the function
(\ref{pin}) is a pure gauge at any region which does not contain the
singular points. However it is  ambiguous in the whole space
therefore globally the f\/ields $ J_k $ are  not a pure gauge.
We recall that the integral $ \oint dz_i \,\hat J_i $ taken along a closed
contour encircling the singularity is not equal to zero, i.e.
depends on the integration contour.

\section{On the calculation of the vacuum energy density.}

The vacuum energy density is  def\/ined  as
\be
\label{ev}
           \varepsilon_{vac}=\frac{1}{4}<0|\theta_{\mu\mu}|0>,
\ee
where $ \theta_{\mu\mu} $   is the trace of the energy-momentum tensor,
the numerical  coeff\/icient is determined by the dimension of the
physical space $  D=4 $.The vacuum energy density is def\/ined as a
sum of the perturbative and non-perturbative contributions.
 Since we are interested in the non-perturbative effects we  will
 now assume that the perturbative contributions into the vacuum
energy density are  subtracted from $ \varepsilon_{vac} $ in eq. (\ref{ev}).
The non-perturbative part of the vacuum energe density is denoted
by $ \varepsilon^n_{vac}$.

It is well known that the dilatation anomaly in gauge theories is
proportional to the  ${\beta}$-functon and the f\/ield strength
squared but the scale independent value of the dilatation anomaly is
only $g^2(F_{\mu\nu}^a)^2 $. Therefore only this part $ \varepsilon^n_{vac}$
has a physical significance. For this reason we shall calculate the vacuum
condensate of $ g^2(F^a_{\mu\nu})^2 $. Notice that  this method seems to
avoid the computation of the dilatation anomaly condensate if
renormalization constants are known at all orders.

Let  the Lagrangian be def\/ined in terms of the renormalisation gauge
f\/ields and constants as
\be
\label{lag}
             L=-\frac{1}{4}\left(F^a_{\mu\nu}\right)^2-\frac{1}{2\alpha}
                 \left(\partial_{\mu}A^a_{\mu}\right)^2-\frac{1}{4}
                  \left(Z_3-1\right)
               \left(\partial_{\mu}A^a_{\nu}-\partial_{\nu}A^a_{\mu}\right)^2.
\ee
The ghost  are omitted because they are essential
only in  perturbative calculations. The gauge f\/ields, the gauge
f\/ixing parameter and the coupling constant are given by
$$
      A_b=Z^{1/2}_3A, \alpha_b=Z_3\alpha,  g_b=Z_gg,
$$
where $ A_b $ ,$ g_b $ ,$ \alpha_b $ are bare quantities.
The quantities $ Z_g $  and $ Z_3 $ are calculated at the one-loop level.
In such a case we have
\be
\label{cz}
           Z_3 = 1+ \frac{g^2_b}{16\pi^2}\left(\frac{13}{6}N-
                        \frac{\alpha_b}{2}\right)\ln M^2/\mu^2,
           Z_g = 1+\frac{g^2_b}{16\pi^2}\frac{13}{6}N\ln M^2/\mu^2,
\ee
where $ \mu^2  $   is a normalization point,\, $ M^2 $ is an ultraviolet
cutoff.

The last term in eq.(\ref{lag}) is a counterterm which is not  gauge
invariant. As above  gauge invariance of the counterterm is not required.

It is known \cite{MS} that the conformal anomaly  $  \theta_{\mu\mu} $
can be obtained as the  variation of  action
when the ultraviolet cutoff  $ M^2 $ changes into  $ (1+\eta)M^2 $
with  the coupling constant  kept fixed.
Here $ \eta $ is the parameter of the global scaling transformation.
Then we can obtain
\be
\label{tm}
              \theta_{\mu\mu}=-\frac{g^2 _b}{64\pi^2}b
                    \left(\partial_{\mu}A^a_{\nu}-\partial_{\nu}
                        A^a_{\mu}\right)^2.
\ee
Here $b=13/6N-\alpha_b/2$.

In terms of the generating functional $ Z_E $ in Euclidian space,\, v.e.v. of
$ \theta{\mu\mu} $ can be written as
\be
\label{gi}
           <0|\theta_{\mu\mu}|0>=
                      Z^{-1}_E\int DA_i\,\theta^R_{\mu\mu}\,exp^{-S_E}
\ee
Here  the quantity $\theta^R_{\mu\mu} $ have been obtained from eq.(\ref{tm})
by  introducing the  regularization.

Let us calculate the vacuum energy density
at some point  $ x_0 $. Let us assume that $ x_0 $ is enclosed by a sphere
$ S^2 $ of a small radius $ r $. To regularize the  quantity
$ \theta_{\mu\mu} $ we define the quantum f\/ields in different
points on the surface of the sphere $ S^2 $. Let us  put
the radius to be  inverse of the parameter $  {\mu} $
from eq.(\ref{mu}).

Notice that making so we spoil the gauge
symmetry.   It  should be restored if the operator is
the gauge invariant quantity. However,since the operator $ \theta_{\mu\mu} $
eq.(\ref{tm}) is not gauge invariant and from the very beginning, we may not
take care of  the gauge symmetry  at all. Also it should
be kept in mind that the quantity $ A^a_4 $ are c-number f\/ields
 and therefore the quantity $(\partial_k A^a_4)^2 $ do not have to be
regularized.
As  will be shown later therefore
the contributions of the electrical strengh
in the non-pertrubative vacuum energy are absent for the static f\/ields.

According to V.N. Gribov \cite{G}, the contributions of the
configurations $ J^a_i  $ can not be compensated by the ghost f\/ields
because the Faddeev-Popov determinat has zero and the standart method
has to be improved.
In the static case the action has  a minimum on this configurations as
 pure gauge configurations in all  physical space besides the point
of the singularity. Here we keep in mind that contributions to the
integral which come from the sphere rigion are ignored,but the contributions
from the surface of the sphere are.

At the begining one may speculate that  the singular point
is  $ x_0 $ where the operator $ \theta_{\mu\mu} $
is defined. We recall that  the point $ x_0 $ is surrounded
by the sphere  of a small radius and area bounded by the sphere
surface is excluded from consideration. In such case the topological
charge equals to zero. Really, the topological charge is
\be
\label{qt}
              Q=\oint_{S_\infty} d\sigma_k\,j_k -\oint_{s_r} d\sigma_k\,j_k=0,
\ee
where $ j_k=\varepsilon_{kij}\partial_iA^a_j\Phi^a $, $ S_\infty $ and
$ s_r $ are the respective surfaces of the sphere at large and small
distances. The result (\ref{qt}) is valid for any infinitely
small quantity $ r $, which has to be considered as a "physical"
zero. The result (\ref{qt}) is explained by
the absence of the singularity in the
region between two sphere surfaces where  the fluctuations are def\/ined.

At first glance it would seem that we discussed a special case
because the singular point coincides with the point in which
the  operator  $ \theta_{\mu\mu} $ is examined. Indeed, this is
not the case. As the function $ \Phi^a(x) $ are subject to
constraint (\ref{cf}) which is independent of a position of
the singular point  and the quantity
$ \left(\partial_i\Phi^a\right)^2 $ is a translational invariant
function therefore  we can give any position to the singularity for
any combination being quadratic
in $ \Phi^a $ or $ \partial_i\Phi^a $. As we shall see,
the v.e.v. of $ \theta_{\mu\mu} $ may be written as one of such combinations
due to  $\theta_{\mu\mu}$ is independent of the position of the singularity.

We recall that the action has the minimum on the f\/ields
$ A^a_i=\frac{1}{g}J^a_i $ therefore v.e.v. of $  \theta_{\mu\mu}  $
can be written as follows
\be
\label{vth}
\ba{c}
 <0|\theta_{\mu\mu}|0>=\lim\limits_{\Delta \to 0}-g^2_bb/64\pi^2
   \left[2\mu^2\left(\partial_i\Phi^a\right)^2
-1/g^2_b\left(\partial_iJ^a_j(x_0-\Delta)\right.\right.-\\
\left.\partial_jJ^a_i(x_0-\Delta)\right)
\left.
\left(\partial_iJ^a_j(x_0+\Delta)-\partial_jJ^a_i(x_0+\Delta)\right)\right].
\ea
\ee
Here, the first term is obtained from the term
$  \left(\partial_iA_4\right)^2=\mu^2\left(\partial_i\Phi\right)^2 $.
This term is the singular one. It can be  verified that the term tends
to infinity as $ 1/r^2 $, when $ r $ goes to zero. To do this one needs to use
parametrization of the function $ \Phi^a(x) =(x-x_0)^a/r $.The quantity
$ 1/r^2 $
may be considered to be of the  order of the ultraviolet cuttoff squared for
a small $  r^2 $. Therefore the first term  gives the contribution to
the perturbative part of the vacuum energy density \cite{NS,K}. Since
we are interested in non-perturbative part of the vacuum energy density
then we shall not discuss the first term.We recall that we assumed that
the perturbative conributions into v.e.v. of $ \theta_{\mu\mu}  $
are subtracted.

The non-perturbative part of the vacuum energy  is associated with
the second term in (\ref{vth}) and can be written as
\be
\label{npth}
\ba{c}
<0|\theta_{\mu\mu}|0> =\lim_{\Delta \to 0}b/64\pi^2
\left(\partial_iJ^a_j(x_0-\Delta) -\partial_jJ^a_i(x_0-\Delta)\right)\\
\left(\partial_iJ^a_j(x_0+\Delta)-\partial_jJ^a_i(x_0+\Delta)\right).
\ea
\ee

It is convinient to rewrite eq.(\ref{npth}) in another form,
making use the eq.(\ref{con})
$$
     F_{ij}=\partial_iJ^a_j - \partial_jJ^a_i + \varepsilon^{abc}J^b_iJ^c_j=0,
$$
and we arrive at
\be
\label{nh}
\ba{c}
<0|\theta_{\mu\mu}|0>=\\
\lim_{\Delta \to 0}b/64\pi^2
\left\{\left(\partial_i\Phi^a(x_0-\Delta)\partial_i\Phi^a(x_+\Delta)\right)^2
\left(\Phi^b(x_0+\Delta)\Phi^b(x_0-\Delta)\right)^2-\right.\\
\left.
-\left(\partial_i\Phi^a(x_0-\Delta)\partial_j\Phi^a(x_0-\Delta)\right)\right.\\
\left.
\left(\partial_i\Phi^b(x_0+\Delta)\partial_j\Phi^b(x_0+\Delta)\right)
\left(\Phi^c(x_0-\Delta)\Phi^c(x_0+\Delta)\right)^2\right\}.
\ea
\ee
Here the terms of the type $ \partial_i\Phi ^a(x_0-\Delta)\Phi^a(x_0+\Delta) $
were omitted because they tend to zero when $ \Delta $ goes to zero.
The quantity $ \partial_i\Phi^a(x_0-\Delta)\partial_j\Phi^a(x_0+\Delta) $
is approximately equal to
$ \delta_{ij}/3\partial_k\Phi^a(x_0-\Delta)\partial_k\Phi^a(x_0+\Delta) $
and due to that we get
\be
\label{svt}
\varepsilon^n_{vac}=\lim_{\Delta \to 0}b/384\pi^2
\left(\partial_i\Phi^a(x_0-\Delta)\partial_i\Phi^a(x_0+\Delta)\right)^2
\left(\Phi^b(x_0-\delta)\Phi^b(x_0+\Delta)\right)
\ee

Now it should be shown that the quantity $ \varepsilon_{vac} $
is scale independent constant. We can do it in two steps. As
the f\/ields $\Phi^a $ have been def\/ined on the surface of the sphere
$ S^2 $ then at the f\/irst step we can place the f\/ields $ \Phi^a(x_0-\Delta)$
and $\Phi^a(x_0+\Delta) $ in one and the same point on the surface keeping
the radius of the sphere constant. At the second step the radius tends to
zero.It  is convenient to  introduce new variables $ \Delta_i $ and
$ r^2=\Delta_i^2 $ holding $ x_0 $ f\/ixed. Making use of the
parametrization of the function $\Phi^a=\gamma\Delta^a/r $
we have
\be
\label{r2}
           \left(\partial_i\Phi^a\right)^2=-\frac{2}{r^2}\gamma^2.
\ee
Here $ \gamma $  is the normalization constant which is given below.

To calculate the quantity $ \Phi^a(x_0-\Delta)\Phi^a(x_0+\Delta) $
one can use eq.(\ref{pin}) and obtain
\be
\label{pex}
  \Phi^a(x_0-\Delta)\Phi^a(x_0+\Delta)=
\Phi^b(x_0)\Phi^c(x_0)\left(P\,\exp\left(-\int\limits_{x_0-\Delta}^{x_0+\Delta}
dz_i\,J_i\right)\right)_{bc}.
\ee
We can write eq.(\ref{pex})  in a more suitable form. To  this end
one expands of exponential function in eq.(\ref{pex}) in powers    of small
$   \Delta $ to second order and  gets that the f\/irst oder
term is equal to zero due to antisymmetric tensor $\varepsilon_{abc} $
and the second order term  contains $\Phi^b\Phi^c $ times $\Phi^b\Phi^c $
and thus one gets $\left(\Phi^2\right)^2 $ =1. Therefore eq.(\ref{pex})
can be written as follows
\be
\label{aa}
\Phi^a(x_0-\Delta)\Phi^a(x_0+\Delta)=\left(P\,\exp\left(-\int\limits_{x_0-
\Delta}^{x_0+\Delta}
dz_i\,J_i\right)\right)_{aa}.
\ee
The contour of integration should be closed  around the sphere with the f\/ixed
radius. Then   substituting (\ref{r2}) and (\ref{aa}) into (\ref{svt})   we
have
\be
\label{evl}
\varepsilon^n_{vac}=\lim_{r \to 0 }b/96\pi^2\left(\frac{\gamma^2}{r^2}
P\,\exp\left(-\oint dz_i\,J_i\right)\right)^2
\ee

To calculate the integral around the circle use is made of the identy
$ \left(J^a_it^a\right)^{bc}=\Phi^b\partial_i\Phi^c - \partial_i\Phi^b\Phi^c  $
which yeilds
\be
\label{ss}
  \oint dz_i\,J_i^{bc} = \oint d\Phi^c\,\Phi^b- d\Phi^b\,\Phi^c=2S^{bc}
\ee
The area $ S^{bc} $ is  enclosed by the circle.The third axis may be oriented
normally to the area $S^{bc} $ in the colour space.The f\/ield $   \Phi^3 $
 is f\/ixed in such a way   that f\/ields $ \Phi^1 $ and $\Phi^2 $
are normalized as $\left(\Phi^i\right)^2=\lambda,  i=1,2 $.
In this case the area $  S^{12} $ equals   $ \pi\lambda $ and the integral
around the circle is $  2\pi\lambda $.
For  $ \varepsilon^n_{vac} $ we have
\be
\label{evf}
 \varepsilon^n_{vac} =b/96\pi^2\left(\frac{\gamma^2}{r^2}\exp(-2\pi\lambda)
\right)^2.
\ee
Here $\varepsilon_{vac} $ is obtained on the scaling mass $\mu^2=1/r^2 $.
Now we should get the vacuum  energy density corresponding to $ r^2=1/M^2 $.
Since the f\/ields were def\/ined on the scaling mass $ \mu^2  $ they
can be   rewritten in terms of the bare  f\/ields. Then we have
$J_{i}=Z^{1/2}_3\,J_i $ and $ \Phi_b=Z^{1/4}\Phi_b $. In this case
the relation between $ \lambda $ and unrenormalized constant
$\lambda_b $ is written as $ \lambda=Z^{-1/2}\lambda_b $. Besides
we have to def\/ine a relation
$ <0|\theta_{\mu\mu}|0>=Z^{-1}_3<0|\theta_{\mu\mu}|0>_b $.
Clearly the constant $ \gamma^2 $ in eq.(\ref{r2}) is $ Z^{-1/2}_3 $,
and we get
\be
\label{em}
\varepsilon^n_{vac}=b/96\pi^2
\left(\mu^2\,\exp\left(-2\pi\lambda_bZ^{-1/2}_3(\mu^2)\right)\right).
\ee
The value $\varepsilon_{vac} $ is scale independent only if
$$
       \mu^2\,\frac{d\varepsilon^n_{vac}}{d\mu^2}=0.
$$
Using  (\ref{em}) and (\ref{cz}) we obtain that if
$$
   \lambda_b=8\pi/(g^2_bb),
$$
then $\varepsilon^n_{vac} $ is scale independent.
Now we  obtain the final result
\be
\label{fr}
 \varepsilon^n_{vac}=b/96\pi^2\,\left(M^2\,
 \exp\left(-16\pi^2/bg^2_b\right)\right)^2.
\ee
The quantity $\varepsilon^n_{vac}$ is scale independent and rises as $ N $
which is agremeant with the familiar result \cite{NS}.

Notice that $\lambda_b$ is a large quantity ($ \lambda\gg1 $)
due to $ g^2_b(M^2)/4\pi\ll~1 $,therefore $\Phi^2_3=1-\lambda_b $
may be negative quantity and this, in its turn, implies that the condition
$ \left(\Phi^a\right)^2=1 $ is not the equation of a sphere in view
of quantum effects, i.e. the monopoles are absent in the scale independent
quantum  theory.

\section{Conclusions.}

In the present paper we have  calculated the non-perturbative part
of the vacuum energy density in pure Y-M theory which is determined by the
static vacuum fluctuations in one-loop level. The result was obtained
by using the physical  ideas which  are derived in the special case of
sigma models \cite{K,V}.

It was shown that the vacuum energy density  in Y-M theory is the scale
independent quantity, which corresponds to the topological charge equal
to zero.

The method of regularization is proposed which is based on the critical
assumption that at each point of the space  in which  a product of
two operators can be  replaced by a sphere $S^2  $ having a small radius
which  is set to zero   at    the very end of the calculation and the
quantum f\/ields are separated in diff\/irent point on the surface of
the sphere. The calculation have been carried out  using functional
integral but can be done by the operator method.
For this aim the vacuum state has been constructed.

Acknowledements.

The author wishes  to thank V.A. Novikov, A.D. Mironov, Yu.M. Makeenko
 for useful discussions.
The paper was partially supported by the  Russian Foundation
of Fundamental Reseach (grant No 98-02-17316).

 \end{document}